\documentclass[aip,apl,10pt,twocolumn,superscriptaddress,10pt]{revtex4-1}
\usepackage{amssymb,amsmath,amsfonts,xcolor,graphicx}
\usepackage{times}
\usepackage{braket}
\usepackage{times}
\usepackage{float}
\usepackage{lipsum}

\begin{document}

\title{\Large  Robust Nonadiabatic Holonomic Quantum Gates on Decoherence-Protected Qubits}

\author{Zhi-Cheng He}
\affiliation{Guangdong Provincial Key Laboratory of Quantum Engineering and Quantum Materials, 
and School of Physics and Telecommunication Engineering, South China Normal University, Guangzhou 510006, China}

\author{Zheng-Yuan Xue}\email{zyxue83@163.com}
\affiliation{Guangdong Provincial Key Laboratory of Quantum Engineering and Quantum Materials, 
and School of Physics and Telecommunication Engineering, South China Normal University, Guangzhou 510006, China}
\affiliation{ Guangdong-Hong Kong Joint Laboratory of Quantum Matter, and Frontier Research Institute for Physics,\\ South China Normal University, Guangzhou 510006, China}


\date{\today}

\begin{abstract}
Obtaining high-fidelity and robust quantum gates is the key  for scalable quantum computation, and one of the promising ways is to  implement quantum gates  using geometric phases, where the influence of local noises can be greatly reduced. To obtain robust quantum gates,  we here propose a scheme for quantum manipulation by combining the geometric phase approach with the dynamical correction technique, where the imperfection control induced \emph{X}-error can be greatly suppressed. Moreover, to be robust against the decoherence effect and the randomized  qubit-frequency shift \emph{Z}-error, our scheme is also proposed based on the polariton qubit, the eigenstates of the light-matter interaction, which is immune to both  errors up to the second order,  due to its near symmetric energy spectrum. Finally, our scheme is implemented on the superconducting circuits, which  also  simplifies  previous implementations. Since  the main errors can be greatly reduced in our proposal,  it provides a promising strategy for scalable solid-state fault-tolerant quantum computation.
\end{abstract}

\maketitle

Quantum computation \cite{qc}, which uses superposition property of quantum states to perform computation, is believed to be in a position that can solve  certain hard computational problems for classical computers. But, during any gate operation, which is the building block of a quantum computer, the noises and decoherence effect will be inevitable, leading to the infidelity of a target quantum gate.  Thus, under the noise and decoherence effects, how to build an advisable quantum system for physical implementation of a quantum computer has attracted much attentions. Among the proposed candidates, superconducting quantum circuits system \cite{sqc4} has being pursued due to its  flexible controllability and easy scalability. Currently, superconducting chip with dozens of transmon qubits \cite{transmon1, transmon2} can be efficiently manipulated to show the quantum advantage \cite{QS2019, QS2021}.

Geometric phases, acquired during a cyclic evolution, was firstly discovered by Berry in the adiabatic process \cite{Abelian}. Then, the Berry phase has been extended to the non-Abelian \cite{non-Abelian} and nonadiabatic   \cite{AA} cases. As geometric phases only depend on the global geometric properties of their evolution paths, they are insensitive to certain local noises. Therefore, geometric phases can naturally find important applications in quantum computation \cite{zanardi, AGQC1,  wxb, ZSL1, UGQCZhu}, where quantum gates are implemented  by using  geometric phases. As the non-Abelian geometric phase is of the matrix form, it can naturally be used to form universal  quantum gates, i.e., the holonomic quantum computation (HQC) \cite{zanardi}. However, its physical implementation is experimentally difficult due to the need of complex interaction between multi-level quantum systems \cite{Duan, cenlx, bjliu}. Using three level system, simplified nonadiabatic HQC (NHQC) has been proposed \cite{NJP, TongDM} and received many renewed theoretical \cite{Singleloop, composite, SingleloopSQ, surface2, eric, Liu18, dd, dd2, Li, zhaopz, wuc, chentime, xu1, xu2, BNHQC, dcg} and experimental \cite{Abdumalikov13, Feng13, Zu14, AC14, nv2017, nv20172, li2017, xuy18, yan2019, zhu2019, yinyi, aimz, aimz2, sunfw1, yuyang, sunfw2, lisai} interests. Recently, further explorations are mainly focus on strengthen  the gate robustness \cite{xu1, Liu18, dd, dd2, Li, wuc, dcg} and further speedup the gates \cite{xu2, zhaopz, chentime, BNHQC}, in experimental accessible setups. However, currently,  this is a still  on-going exploration.

Here, we propose a scheme to implement NHQC, on decoherence-protected polariton qubits, with the dynamically correction technique  (DCT) \cite{dc1, dc2, dc3}.  The polariton qubits is formed by the eigenstates of 
the Jaynes-Cummings (JC) Hamiltonian, which can be implemented in a typical circuit QED setup \cite{cqed}. Meanwhile,  the two eigenstates of the single-excitation subspace are chosen as our logical qubit-states, which are decoherence insensitive. Moreover,  besides the decoherence effect,  the control induced \emph{Z}-error  in  superconducting  circuits mainly originates from randomized  qubit-frequency drifts, which vary in a time scale that is much longer than the gate-time, and thus can be treated as a constant during a quantum gate. Remarkably, due to the near symmetric spectrum of the JC Hamiltonian, our logical qubit states are immune to the static \emph{Z}-error up to the second order. Note that, this merit is not shared by previous schemes based on the dressed-state qubits \cite{xue, wangym1, wangym}, 
as they have not used the dressed states from the same excitation subspace as logical qubit-states, and the dephasing protection there can only be achieved by further increasing the circuit complexity \cite{wangym}. From the implementation point of view, this is another merit of our scheme, as it is based on simplified setups and without auxiliaries as in Refs.  \cite{xue, wangym1, wangym}. In addition, in order to be robust against the control induced \emph{X}-error, from the imperfection control of the amplitude of the driving field and thus can be time-dependent during a quantum gate, we also incorporate the DCT. We  show that this correction is better suitable for our polariton qubit than the conventional bare qubits.  Note that, decoherence protection can also be obtained in  Refs. \cite{Polariton}, but the control-error insensitivity is not shared there. Besides, only the implementation of single-qubit gates are considered there. Therefore,  our scheme provides a promising way to achieve fault-tolerant and scalable solid-state quantum computation.



\begin{figure}[tbp]
  \centering
  \includegraphics[width=0.95\linewidth]{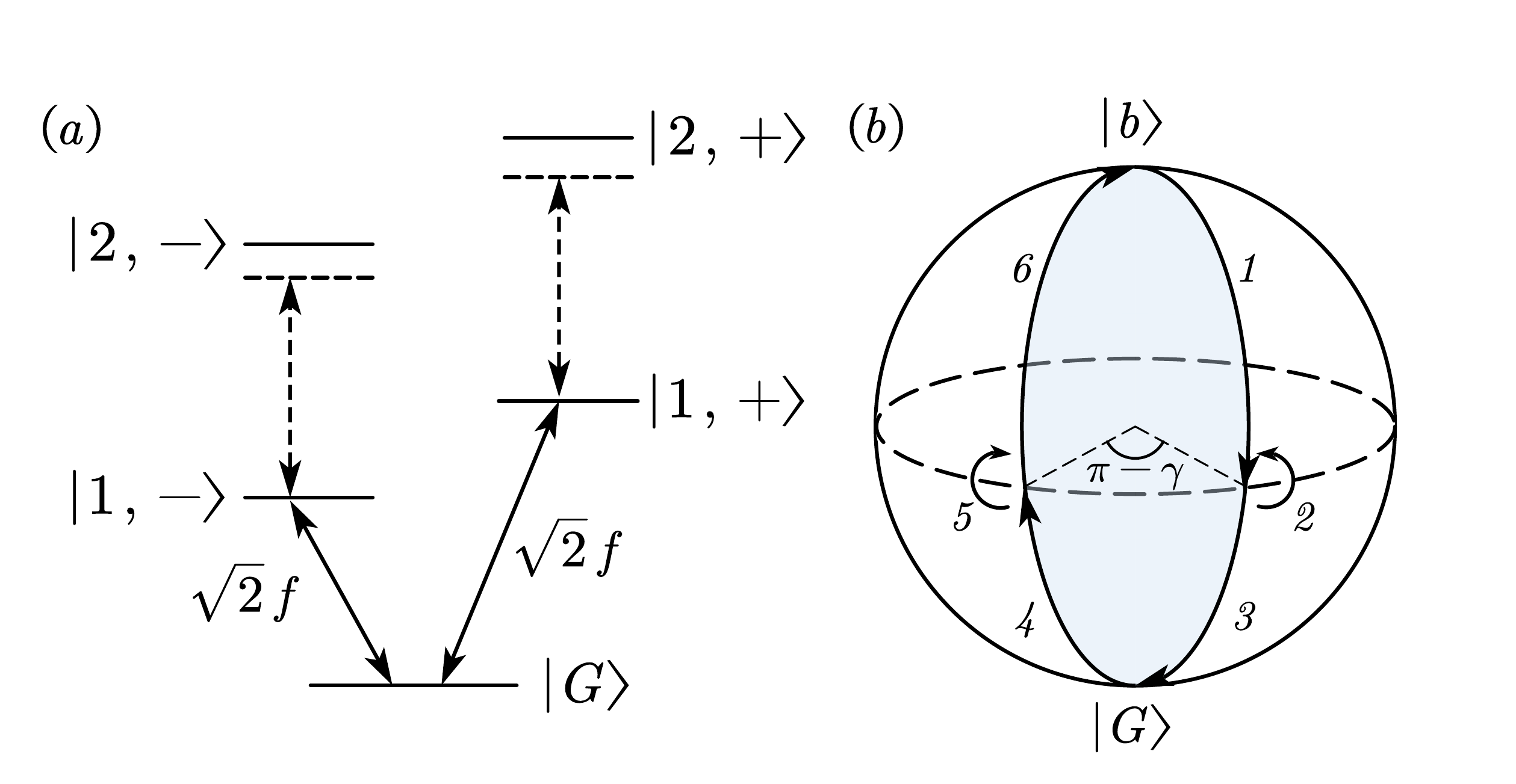}
  \caption{Illustration of our gate construction. (a) The spectrum of the JC Hamiltonian. By setting appropriate frequency of an external driving microwave field, the transitions between  $\ket{G}$ and $\ket{1,\pm}$ states can be addressed, forming a V-configuration artificial atom. (b) The process to obtain a pure geometric phase in the $\ket{b}$ state, where the Bloch sphere representation is shown within the subspace spanned by $\{\ket{b},\ket{G}\}$, and the dynamical corrections are inserted at the middle of the two longitude lines. }
  \label{Bloch_Spectrum}
\end{figure}

Firstly, we present the implementation of our  polariton qubit and show its decoherence-protection merit. Considering that a transmon qubit  is capacitively coupled to a microwave cavity, e.g.,  in a conventional circuit QED setup \cite{cqed}. Assuming $\hbar = 1$ hereafter, the interaction Hamiltonian is of the well-known  JC model form as
\begin{eqnarray}
\label{HJC}
H_{JC}=\frac{\omega_q}{2}\sigma_{z} + \omega_c a^{\dagger} a + g(a \sigma^{+} + a^{\dagger} \sigma^{-})
\end{eqnarray}
where $\omega_q$ and $\omega_c$ represent the frequencies of the qubit and cavity,  $\sigma_{z}$ is the Pauli \emph{Z} matrix, $\sigma^{-}=\ket{0}_q \bra{1}$ with $\ket{0}_q$ and $\ket{1}_q$ are the ground  and  first excited states of the transmon qubit, $g$ is the qubit-cavity interaction strength. Label $\ket{n}_{c}$ as the Fock space of the cavity and $\ket{0,n}=\ket{0}_q \ket{n}_c$ as the product states of the qubit and cavity. The eigenvalues of the Hamiltonian in Eq. (\ref{HJC}) can be calculated
\begin{eqnarray}
\label{Eigenvalues}
E_{n,\pm}=n \omega_c +  ( \delta \pm \sqrt{\delta^{2}+4ng^{2}})/2,
\end{eqnarray}
where $\delta=\omega_c-\omega_q$ is the cavity-qubit detuning, and, except the ground state $\ket{G}=\ket{0, 0}$, other eigenstates are grouped as
\begin{equation}
\label{Eigenstates}
\ket{n,\pm}=\cos{\alpha_n}\ket{0,n} \pm \sin{\alpha_n}\ket{1,n-1},
\end{equation}
where $\alpha_n=2g\sqrt{n+1}/\delta$.
In the subspace spanned by three lowest eigenstates, i.e., the zero- and single-excitation subspace $S_1=\{\ket{G}, \ket{1, \pm} \}$,   transitions between $\ket{G}$ and $\ket{1,\pm}$ can be implemented by  microwave fields with proper frequency addressing, and thus they form a V-configuration artificial atom, as shown in Fig. \ref{Bloch_Spectrum}(a).

The dressed states in the single-excitation subspace $\ket{1,\pm}$  are chosen as our logic qubit states,  denoted by   $\ket{\pm}$  for   simplicity hereafter, which is qubit-decoherence insensitive. As the lifetime of a cavity \cite{cavity} may be much longer than that of a transmon qubit \cite{sqc4}, we here only consider the decoherence from the transmon-qubit system. For the decoherence noise in the form of $h_x \sigma_x$, due to its low-frequency nature, it can only drive the the transitions between a qubit state to the ground state, in a very large detuned way. Up to the second order, the noise induced shift  of the eigenenergies  are $\delta E_\pm ^x \approx \mp(h_x/\omega_q)^2 \times g \ll h_x$, where the superscript $x$ indicates the shift is originated from the $\sigma_x$ noise and the subscripts $\pm$ denote the shifts are for the logical qubit-states $\ket{\pm}$, respectively. Similarly, for the $h_z \sigma_z$ noise, the induced  eigenenergy shifts are $\delta E_\pm ^z \approx \pm (h_z/g)^2 \times g/2 \ll h_z$. That is, for our polariton qubit, the noise effects from the decoherence of the transmon  are suppressed to the second order, instead of the leading first order in the conventional case, where the energy shifts will be on the order of $h_{x/z}$ for the $\sigma_{x/z}$ noise.


Now, we proceed to present the holonomic manipulation on a polariton logical qubit,  which can be achieved when suitable microwave drives applying on the transmon. In the subspace $S_1$, $H_{JC}$ will be a diagonalized matrix with the elements being their eigenvalues, i.e.,
\begin{eqnarray}
\label{H0}
H_0=
\left(
\begin{array}{ccc}

E_{0}         &0             &0 \\
0           &E_{1, -}         &0\\
0           &0             &E_{1, +}\\

\end{array}
\right).
\end{eqnarray}
To manipulate the states in $S_1$, we consider applying a driven field on the transmon in the form of $H_d=\sqrt{2}f(t)\sigma_{x}$. Here, to induce both the transitions between $\ket{G}$ and $\ket{\pm}$, the driven field can be chosen as
\begin{eqnarray}
f(t)=\Omega_1(t) \cos{(\omega_1 t-\varphi_1)}+\Omega_2(t) \cos{(\omega_2 t-\varphi_2)},
\end{eqnarray}
where $\Omega_n$, $\omega_n$ and $\varphi_n$, with $n=1,2$, is the amplitude, frequency and phase of the $n$th tone of the driven field, respectively. As $\bra{G}\sigma_{x}^{q}\ket{\pm}=\bra{\pm}\sigma_{x}^{q}\ket{G}=\pm 1/\sqrt{2}$,
\begin{eqnarray}
\label{Driven}
H_{d}= f(t) (\ket{G}\bra{+}  -  \ket{G}\bra{-})+\text{H.c.}.
\end{eqnarray}
In order to investigate the qubit dynamics in a holonomic way, the driven field is set as $\omega_{1}=\omega_{-}$ and $\omega_{2}=\omega_{+}$, with $\omega_{\pm}=E_{1, \pm}-E_{0}$ being the transition frequencies between the corresponding states. Thus, in the interaction picture with respect to $H_0$, the interaction Hamiltonian $H_d$  will be
\begin{align}
\label{H holonomic}
H_{1}
=\frac{\Omega(t)}{2}e^{i\varphi_2}\left(\cos{\frac{\theta}{2}}\ket{+}-\sin{\frac{\theta}{2}} e^{i\varphi}\ket{-}\right)\bra{G} +\text{H.c.},
\end{align}
where $\Omega(t)=\sqrt{\Omega_{1}(t)^2+\Omega_{2}(t)^2}$, $\tan{(\theta/2)}=\Omega_1(t)/\Omega_2(t)$ with $\theta$ being a constant, and $\varphi=\varphi_1-\varphi_2$. Here, to get the above interaction, we have set $ \omega_{n} \gg g \gg\Omega_{n}$, so that the rotating wave approximation (RWA) can be justified. In this way, the transitions within the artificial atom can be induced, which exhibits the bright and dark states of
\begin{subequations}
\label{Eigenstates}
\begin{align}
&\ket{b}=\cos{\frac{\theta}{2}} \ket{+}-\sin{\frac{\theta}{2}} e^{i\varphi}\ket{-},\\
&\ket{d}=\sin{\frac{\theta}{2}} e^{-i\varphi}\ket{+}+\cos{\frac{\theta}{2}}\ket{-}.
\end{align}
\end{subequations}
As the dark state $\ket{d}$ is decoupled, the quantum dynamics under Eq. (\ref{H holonomic}) is captured by the resonant coupling between $\ket{b}$ and $\ket{G}$ states. When the cyclic evolution condition $\int_{0}^{T} \Omega(t) dt=2\pi$ is met, there is no population in state $\ket{G}$, and thus only the bright state $\ket{b}$  acquires a geometric phase factor $(\pi-\gamma)$. Consequently,  one can obtain a quantum gate in the subspace spanned by $\ket{\pm}$ depending on the parameters of $\theta$  and $\varphi$.

\begin{figure}[tbp]
  \centering
  \includegraphics[width=0.95\linewidth]{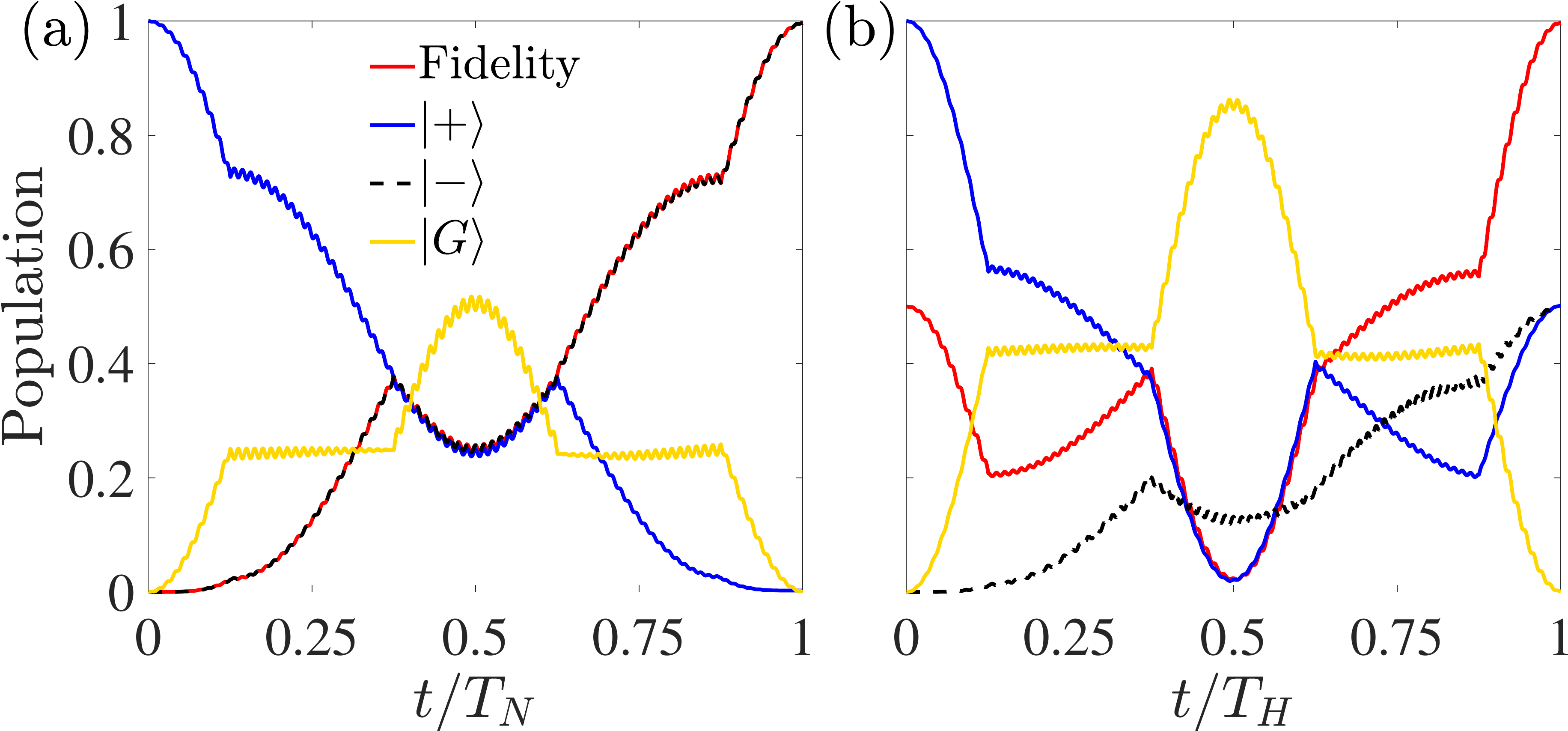}
  \caption{State population and the corresponding fidelity dynamics of holonomic single-qubit (a)  NOT   and (b) Hadamard gates with the initial state being $|+\rangle$.}
  \label{Population_X&H}
\end{figure}

To achieve a fast universal set of single-qubit holonomic gates, single-loop way has been widely applied \cite{Singleloop,SingleloopSQ}. Here, on this base, the DCT will also be  applied, and then the total cyclic evolution consists of six pulses with different driving phases, i.e.,
\begin{subequations}
\begin{align}
\label{Phase of driven field}
&\int^{T_1}_{0}\frac{\Omega(t)}{2}dt=\frac{\pi}{4}, \varphi_1=\phi+\pi, \varphi_2=\pi,\\
&\int^{T_2}_{T_1}\frac{\Omega(t)}{2}dt=\frac{\pi}{2}, \varphi_1=\phi+\pi+\frac{\pi}{2}, \varphi_2=\pi+\frac{\pi}{2},\\
&\int^{T_3}_{T_2}\frac{\Omega(t)}{2}dt=\frac{\pi}{4}, \varphi_1=\phi+\pi, \varphi_2=\pi,\\
&\int^{T_4}_{T_3}\frac{\Omega(t)}{2}dt=\frac{\pi}{4}, \varphi_1=\phi+\gamma, \varphi_2=\gamma,\\
&\int^{T_5}_{T_4}\frac{\Omega(t)}{2}dt=\frac{\pi}{2}, \varphi_1=\phi+\gamma+\frac{\pi}{2}, \varphi_2=\gamma+\frac{\pi}{2},\\
&\int^{T_6}_{T_5}\frac{\Omega(t)}{2}dt=\frac{\pi}{4}, \varphi_1=\phi+\gamma, \varphi_2=\gamma.
\end{align}
\end{subequations}
The geometric path of this evolution process in the subspace \{$\ket{b}$,$\ket{G}$\} is illustrated  in Fig. \ref{Bloch_Spectrum}(b). Finally, under the cyclic evolution condition, a certain pure geometric phase can be obtained in the $\ket{b}$,  which leads to a holonomic gate in the qubit-state space $\{\ket{-},\ket{+}\}$ as
\begin{eqnarray}
\label{Gate}
U&&=
\nonumber\left(
\begin{array}{cc}
\cos{\frac{\gamma}{2}}-i\sin{\frac{\gamma}{2}}\cos{\theta}       &-i\sin{\frac{\gamma}{2}}\sin{\theta} e^{i\varphi}\\
-i\sin{\frac{\gamma}{2}}\sin{\theta} e^{-i\varphi}               &\cos{\frac{\gamma}{2}}+i\sin{\frac{\gamma}{2}}\cos\theta
\end{array}
\right)\\
&&=\exp{(-i\frac{\gamma}{2} \bm{n}\cdot \bm{\sigma)}}
\end{eqnarray}
which describes a rotation operation around the axis $\bm{n}=(\sin{\theta}\cos{\varphi},\sin{\theta}\sin{\varphi},\cos{\theta})$ by a angle $\gamma$, thus a universal single-qubit holonomic gate can be obtained.
Since the parallel-transport condition $\bra{b(t)}H\ket{d(t)}=0$ is always satisfied during the evolution and the cyclic condition is also met, $\gamma$ is of the geometric nature.


The performance of our holonomic gate can be evaluated by considering the influence of dissipation using the the  Lindblad master equation
\begin{eqnarray}
\label{Driven field}
\dot{\rho}=i[\rho,H]+\frac{\kappa}{2}\mathcal{L}(a)+\frac{\Gamma_1}{2}\mathcal{L}(\sigma^{-}) +\frac{\Gamma_2}{2}\mathcal{L}(\sigma_{z}),
\end{eqnarray}
where $\rho$ is the density matrix of the system, $H=H_0+H_d$, $\mathcal{L}(A)=2A\rho A^{\dagger}- A^{\dagger}A\rho-\rho A^{\dagger}A$ is the Lindblad operator, $\kappa$, $\Gamma_{1}$ and $\Gamma_{2}$ are decay rate of the cavity, decay and dephasing rate of the qubits, respectively. For example, a NOT and Hadamard gates can be realized by setting $(\gamma, \theta, \varphi)=(1, 1/2, 0)\pi$  and $(\gamma, \theta, \varphi)=(1, 1/4, 0)\pi$, respectively. Considering the limitation of the rotating wave approximation, the coupling strength  is set as $g=\omega_c/20$ with  the cavity frequency being $\omega_c = 2\pi \times 8$ GHz, and $\Omega(t)= \Omega_0=g/20 $. Within current technique, we set   $\kappa=2\pi \times 0.1$ kHz \cite{cavity}, $\Gamma_1=\Gamma_2=\Gamma=2\pi \times 4$ kHz \cite{sqc4}. Assuming the initial state being $|+\rangle$, in Fig. \ref{Population_X&H}, we plot the quantum state population and the corresponding fidelity dynamics for holonomic  NOT   and  Hadamard gates, with the state-fidelity being $99.58\%$ and $99.68\%$, respectively. Furthermore, average over the 2211 possible initial states of $|\psi\rangle_i=\cos\theta' |-\rangle + \sin\theta' e^{i\varphi '}|+\rangle$ with uniform distributed $\theta'$ and $\varphi'$, and the different value are set to be 201 and 11, respectively; as the gate fidelity is  sensitive to $\theta'$ and insensitive to $\varphi'$. Then, the gate-fidelity \cite{gatefidelity} for the NOT and Hadamard gates can be obtained as  $99.74\%$ and $99.63\%$, respectively.

\begin{figure}[tbp]
  \centering
  \includegraphics[width=0.95\linewidth]{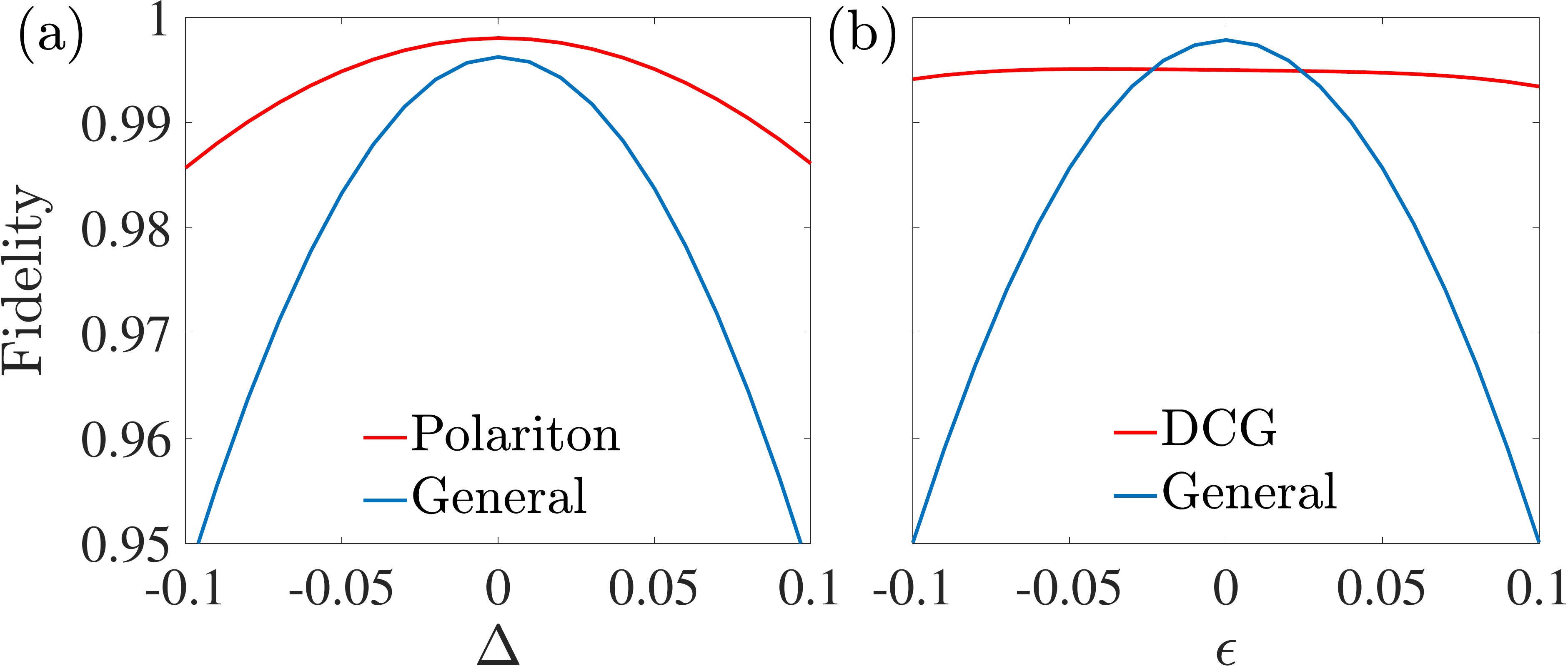}
  \caption{Comparison of the gate performance for the single-qubit Hadamard gate  with different strategies. Both errors are defined with respect to the corresponding driving amplitudes using  error fractions. (a) Our  scheme is more robust than the general   single-loop NHQC schemes without encoding for \emph{Z}-error. (b) With the DCT, our scheme can be much robust than the conventional  scheme for \emph{X}-error. }
  \label{X_Z_error}
\end{figure}

We further demonstrate the robustness of our scheme against the qubit-frequency shift \emph{Z}-error and driving amplitude \emph{X}-error, which will introduce additional term in the interaction Hamiltonian in the form of $\Delta\Omega_0 \sigma_z /2$   and $\epsilon\Omega(t)\sigma_x$, respectively, with $\Delta$ and $\epsilon$ being the error fractions. As discussed before,  the \emph{Z}-error is assumed to be time-independent and  the \emph{X}-error  can be time-dependent, during a holonomic gate on superconducting circuits. We numerically compare our scheme to general NHQC scheme in both error cases. For the \emph{Z}-error, as shown in Fig. \ref{X_Z_error}(a), our scheme is more robust than the single-loop NHQC scheme \cite{Singleloop, SingleloopSQ} without encoding. To fight-against the time-dependent \emph{X}-error, we compare our scheme with DCT and choose $\Omega(t)=\Omega_0 \sin^2(\pi t/T)$ for all pulses, the enhancement of the gate robustness in this case is shown in  Fig. \ref{X_Z_error}(b).  As a result, the comparisons  show  that our scheme is more robust against both  errors. We also note that, with the DCT, the gate-time  will be prolonged, which will introduce more decoherence induced gate error, and thus there has a crosscut for the two curves in  Fig. \ref{X_Z_error}(b). However, with the increasing of the coherent times  of the superconducting quantum circuit \cite{sqc4}, this effect will be negligible small.


Next, we turn to the implementation of nontrivial two-qubit gates in a similar holonomic way, with two polariton qubits being coupled through their corresponding cavities, which simplified the previous implementations \cite{xue, wangym1, wangym}, as an auxiliary logical-qubit from the JC coupling is needed to connect two data-logical-qubits there. In general case, the inter-cavity coupling strength $J(t)$ can be in a time-dependent form \cite{tunephoto}, and the Hamiltonian of this system is
\begin{eqnarray}
\label{H of two qubit}
H_{\text{coup}}=H_{l}+H_{r}+J(t)(a_{l}a_{r}^{\dagger}+a_{l}^{\dagger}a_{r})
\end{eqnarray}
where the $H_{l}$ and $H_{r}$ denote  the free Hamiltonian in Eq. (\ref{H0}) for the left and right polariton qubits, $a_{l/r}^{\dagger}$ and $a_{l/r}$ denote the creation and annihilation operator for the left/right cavities.
Corresponding to the single-qubit notation, our two-qubit gate need to be achieved in the subspace $S_2=\{\ket{++}, \ket{+-}, \ket{-+}, \ket{--}\}$ with $\ket{--}=\ket{-}_l\otimes\ket{-}_r$. Due to the  exchange interaction of photons, the transition between two-qubit subspace and double excited subspace $\{\ket{2-,G}, \ket{2+,G}, \ket{G,2-}, \ket{G,2+}\}$ can be driven, where $\ket{2-,G}=\ket{2,-}_l\otimes\ket{G}_r$. Thus, by setting an appropriate time-dependent function of $J(t)$, only desired transitions can be induced.

\begin{figure}[tbp]
  \centering
  \includegraphics[width=0.95\linewidth]{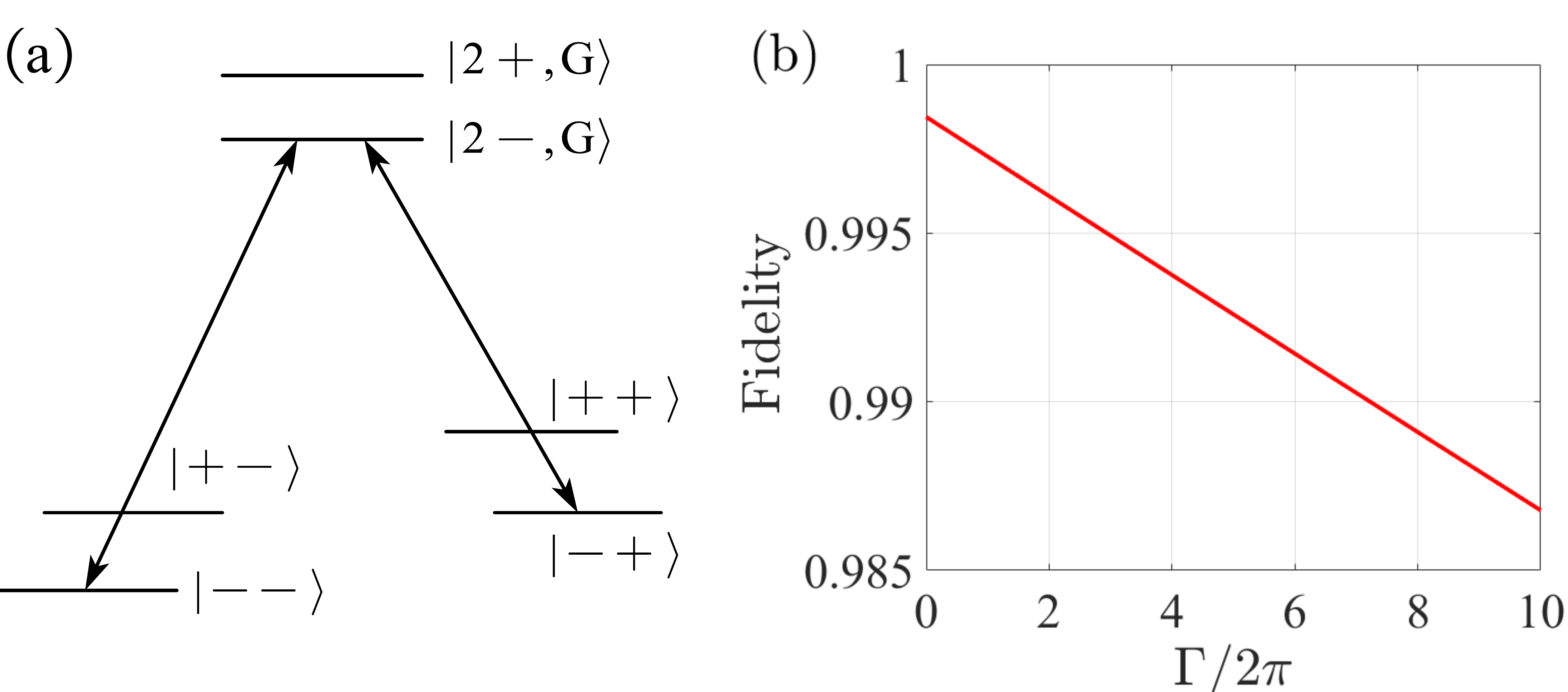}
  \caption{Implementing of  nontrivial holonomic two-qubit gates. (a) The target coupling configuration.  By choosing appropriate parameters, the energy level splitting could be large enough to meet the RWA. (b) The fidelity of CNOT gate as a function of the qubit-decoherence rate $\Gamma$. }
  \label{figuretwo}
\end{figure}

As show in Fig. \ref{figuretwo}(a), we  implement the gate by using an ancillary state $\ket{2-,G}$ in the subspace $S_a=\{\ket{2-,G},\ket{-\pm}\}$.  As $\bra{-\pm}a_{1}a_{2}^{\dagger}\ket{2-,G}=1/\sqrt{2}$, the inter-cavity coupling, in the interaction picture with respect to the free Hamiltonian of $H_l+H_r$,  is
\begin{eqnarray}
\label{H twobit}
H_{\text{I}}= {1 \over \sqrt{2}}
\left(
\begin{array}{ccc}
0                        &J(t)e^{-i\omega_{-}'t}    &J(t)e^{-i\omega_{+}'t}\\
J(t)e^{i\omega_{-}'t}     &0                        &0\\
J(t)e^{i\omega_{+}'t}     &0                        &0
\end{array}
\right),
\end{eqnarray}
where $\omega_{\pm}'=E_{2-,G}-E_{-\pm}$. When setting $J(t)=\sqrt{2}[J_{1}(t) \cos{(\omega_{1}'t-\phi_1)}+J_{2}(t) \cos{(\omega_{2}'t-\phi_2})]$, choosing  $\omega_{1}'=\omega_{-}'$, $\omega_{2}'=\omega_{+}'$, and letting  $\omega_{\pm}'\gg \{J_1, J_2\}$ so that the RWA is met, the above  Hamiltonian will reduce to
\begin{small}\begin{align}
H_{2} =\frac{J_{c}(t)}{2}e^{i\phi_2}\left(\cos{\frac{\theta}{2}}\ket{-+}+\sin{\frac{\theta}{2}} e^{i\phi}\ket{--}\right)\bra{2-,G}+\text{H.c.},
\end{align}\end{small}
where $J_{c}(t)=\sqrt{J_{1}(t)^2+J_{2}(t)^2}$, $\tan{(\vartheta/2)}=J_{1}(t)/J_{2}(t)$ with $\vartheta$ being a constant, and $\phi=\phi_1-\phi_2+\pi$.
Besides, the leakage to other double-excitation subspace may also be occur,  due to the oscillating nature of $J(t)$, e.g., $\bra{+-}a_{1}a_{2}^{\dagger}\ket{2+,G}\neq 0$, which can also be suppressed by the RWA. Thus the harmful transitions will become high frequency oscillation term, which can be removed by RWA. In addition, the manipulation with DCT can  still be useful to fight against X-error here, but it will  twice the gate-time, which will lead to larger gate-infidelity comparing with the singe-qubit gate case. So, in the two-qubit gate cases, we consider the implementation of the single-loop holonomic gates.

Eq. (\ref{H twobit}) has the same coupling configuration as Eq. (\ref{H holonomic}), and thus holonomic quantum manipulation can be induced in $S_a$  similar to the single-qubit  case. When  $J_c$ is chosen to meet
\begin{subequations}
\begin{align}
\label{Phase of coupling strength}
&\int^{T_1}_{0} J_{c}(t)dt= \pi, \phi_1=\phi, \phi_2=\pi,\\
&\int^{T_2}_{T_1} J_{c}(t)dt=\pi, \phi_1=\phi+\alpha-\pi, \phi_2=\alpha,
\end{align}
\end{subequations}
and $\int^{T}_{0}J_c(t)dt=2\pi$ to form a cyclic evolution, the universal nontrivial two-qubit holonomic gate in $S_2$ can be obtained as
\begin{eqnarray}
\label{twobit gate}
U_2=\left( \begin{array}{cccc}
1   &0   &0   &0\\
0   &1   &0   &0\\
0     &0    & \cos{\frac{\alpha}{2}}-i\sin{\frac{\alpha}{2}}\cos{\vartheta}  &-i\sin{\frac{\alpha}{2}}\sin{\vartheta} e^{i\phi}     \\
0     &0    &  -i\sin{\frac{\alpha}{2}}\sin{\vartheta} e^{-i\phi}          &\cos{\frac{\alpha}{2}}+i\sin{\frac{\alpha}{2}}\cos\vartheta
\end{array}\right).\notag\\
\end{eqnarray}

Now as a typical example, by setting the parameters of $\alpha=\pi$, $\vartheta=\pi/2$ and $\phi=0$, once can implement the  CNOT gate. The devices are set up as $\omega_{ql}=2\pi\times 7.8$ GHz, $\omega_{qr}=2\pi\times 4.7$ GHz and $g_{l/r}=\omega_{ql/r}/20$ for the left and right polariton qubit. The inter-cavity coupling strengths are chosen as $J_{1}=J_{2}=2\pi\times 5$ MHz to meet the RWA. The fidelity of this CNOT gate can be $99.37\%$ with the decoherence of $\kappa=2\pi\times 0.1$ kHz and $\Gamma=2\pi\times 4$ kHz for both polariton qubits. Furthermore, we plot the CNOT gate fidelity with different decoherence rates of the transmon-qubit, shown in Fig. \ref{figuretwo}(b), which shows that as the decrease of the decoherence rates, holonomic gates with DCT and other  optimization strategies can be more and more promising.

In summary, we propose a scheme to implement non-adiabatic holonomic quantum gates on polariton qubits. We have numerically demonstrated that our scheme is more robust compare to the general NHQC case in terms of robust against both \emph{Z}- and \emph{X}-errors. Remarkably, this enhancement does not increase the circuit complexity  and without auxiliaries, and thus simplifies previous explorations.  Therefore, it provides a promising strategy toward robust and scalable solid-state quantum computation.

\bigskip
\noindent {\bf Data Availability Statement}\\
The data that support the findings of this study are available from the corresponding author upon reasonable request.

\noindent{\bf Acknowledgements}\\
We thank Dr Yimin Wang for helpful suggestions. This work was supported by
the National Natural Science Foundation of China (Grant No. 11874156), and Science and Technology Program of Guangzhou (Grant No. 2019050001).


\begin{thebibliography}{99}

{\small

\bibitem{qc}
M. A. Nielsen and I. L. Chuang,
\emph{Quantum Computation and Quantum Information} (Cambridge University Press, 2000).



\bibitem{sqc4}
M. H. Devoret and R. J. Schoelkopf,
Science \textbf{339}, 1169 (2013).


\bibitem{transmon1}
J. Koch, T. M. Yu, J. Gambetta, A. A. Houck, D. I. Schuster, J. Majer, A. Blais, M. H. Devoret, S. M. Girvin, and R. J. Schoelkopf,
Phys. Rev. A \textbf{76}, 042319 (2007).

\bibitem{transmon2}
J. Q. You, X. Hu, S. Ashhab, and F. Nori,
Phys. Rev. B \textbf{75}, 140515(R) (2007).

\bibitem{QS2019}
F. Arute, K. Arya, R. Babbush, \emph{et al}.,
Nature (London) \textbf{574}, 505 (2019).

\bibitem{QS2021} Y. Wu, W.-S. Bao, S. Cao, \emph{et al}.,
arXiv:2106.14734.


\bibitem{Abelian}
M. V. Berry,
Proc. R. Soc. London A \textbf{392}, 45 (1984).

\bibitem{non-Abelian}
F. Wilczek and A. Zee,
Phys. Rev. Lett. \textbf{52}, 2111 (1984).

\bibitem{AA}
Y. Aharonov and J. Anandan,
Phys. Rev. Lett. \textbf{58}, 1593 (1987).

\bibitem{zanardi}
P. Zanardi and M. Rasetti,
Phys. Lett. A \textbf{264}, 94 (1999).

\bibitem{AGQC1}
J. Pachos, P. Zanardi, and M. Rasetti,
Phys. Rev. A \textbf{61}, 010305(R) (1999).


\bibitem{wxb}
X.-B. Wang and M. Keiji,
Phys. Rev. Lett. \textbf{87}, 097901 (2001).

\bibitem{ZSL1}
S.-L. Zhu and Z. D. Wang,
Phys. Rev. Lett. \textbf{89}, 097902 (2002).

\bibitem{UGQCZhu}
S.-L. Zhu and Z. D. Wang,
Phys. Rev. Lett. \textbf{91}, 187902 (2003).


\bibitem{Duan}
L.-M. Duan, J. I. Cirac, and P. Zoller,
Science \textbf{292}, 1695 (2001).

\bibitem{cenlx} L. X. Cen, X. Q. Li, Y. J. Yan, H. Z. Zheng, and S. J. Wang,
Phys. Rev. Lett. {\bf 90}, 147902 (2003).

\bibitem{bjliu}
B.-J. Liu, Z.-H. Huang, Z.-Y. Xue, and X.-D. Zhang,
 Phys. Rev. A \textbf{95}, 062308 (2017).

\bibitem{NJP}
E. Sj\"{o}qvist, D. M. Tong, L. M. Andersson, B. Hessmo, M. Johansson, and K. Singh,
New J. Phys. \textbf{14}, 103035 (2012).

\bibitem{TongDM}
G. F. Xu, J. Zhang, D. M. Tong, E. Sj\"{o}qvist, and L. C. Kwek,
Phys. Rev. Lett. \textbf{109}, 170501 (2012).


\bibitem{Singleloop}
E.~Herterich and E.~Sj\"{o}qvist,
Phys. Rev. A \textbf{94}, 052310 (2016).


\bibitem{composite} G. F. Xu, P. Z. Zhao, T. H. Xing, E. Sj\"{o}qvist, and D. M. Tong,
Phys. Rev. A \textbf{95}, 032311 (2017).

\bibitem{SingleloopSQ}
Z.-P. Hong, B.-J. Liu, J.-Q. Cai, X.-D. Zhang, Y. Hu, Z. D. Wang, and Z.-Y. Xue,
Phys. Rev. A \textbf{97}, 022332 (2018).


\bibitem{surface2} J. Zhang, S. J. Devitt, J. Q. You, and F. Nori,
Phys. Rev. A \textbf{97}, 022335 (2018).

\bibitem{eric} N. Ramberg and E. Sj\"{o}qvist,
Phys. Rev. Lett. {\bf 122}, 140501 (2019).

\bibitem{xu1}
G. Xu and G. Long
Phys. Rev. A \textbf{90}, 022323


\bibitem{Liu18}
B.-J. Liu, X.-K. Song, Z.-Y. Xue, X. Wang, and M.-H. Yung,
Phys. Rev. Lett. \textbf{123}, 100501 (2019).




\bibitem{dd} Y. Sekiguchi, Y. Komura, and H. Kosaka,
Phys. Rev. Appl. {\bf 12}, 051001 (2019).

\bibitem{dd2}  X. Wu and P. Z. Zhao,
Phys. Rev. A {\bf 102}, 032627 (2020).

\bibitem{Li}
S. Li, T. Chen, and Z.-Y. Xue,
Adv. Quantum Technol. \textbf{3}, 2000001 (2020).

\bibitem{wuc}
 C. Wu, Y. Wang, X.-L. Feng, and J.-L. Chen,
 Phys. Rev. Appl. \textbf{13}, 014055 (2020).

\bibitem{dcg} S. Li and Z.-Y. Xue,
arXiv:2012.09034


\bibitem{xu2}
G. F. Xu, D. M. Tong, and E. Sj\"{o}qvist,
Phys. Rev. A \textbf{98}, 052315

\bibitem{zhaopz} P. Z. Zhao, K. Z. Li, G. F. Xu, and D. M. Tong,
Phys. Rev. A  {\bf 101}, 062306 (2020).

\bibitem{chentime}
T. Chen, P. Shen, and Z.-Y. Xue,
Phys. Rev. Appl. {\bf 14}, 034038 (2020).



\bibitem{BNHQC}
B.-J. Liu, Z.-Y. Xue, and M.-H. Yung,
arXiv:2001.05182.



\bibitem{Abdumalikov13}
A.~A. Abdumalikov, J.~M. Fink, K.~Juliusson, M.~Pechal, S.~Berger, A.~Wallraff, and S.~Filipp,
Nature (London) \textbf{496}, 482 (2013). 

\bibitem{Feng13}
G.~Feng, G.~Xu, and G.~Long,
Phys. Rev. Lett. \textbf{110}, 190501 (2013).

\bibitem{Zu14}
C.~Zu, W.-B. Wang, L.~He, W.-G. Zhang, C.-Y. Dai, F.~Wang, and L.-M. Duan,
Nature (London) \textbf{514}, 72 (2014).   

\bibitem{AC14}
S. Arroyo-Camejo, A. Lazariev, S. W. Hell, and G. Balasubramanian,
Nat. Commun. \textbf{5}, 4870 (2014).

\bibitem{nv2017}
Y. Sekiguchi, N. Niikura, R. Kuroiwa, H. Kano, and H. Kosaka,
Nat. Photonics \textbf{11}, 309 (2017).

\bibitem{nv20172}
B. B. Zhou, P. C. Jerger, V. O. Shkolnikov, F. J. Heremans, G. Burkard, and D. D. Awschalom,
Phys. Rev. Lett. \textbf{119}, 140503 (2017).

\bibitem{li2017}
H. Li, L. Yang, and G. Long,
Sci. China: Phys., Mech. Astron. \textbf{60}, 080311 (2017).

\bibitem{xuy18}
Y. Xu, W. Cai, Y. Ma, X. Mu, L. Hu, T. Chen, H. Wang, Y. P. Song, Z.-Y. Xue, Z.-Q. Yin, and L. Sun,
Phys. Rev. Lett. \textbf{121}, 110501 (2018).

\bibitem{yan2019}
T. Yan, B.-J. Liu, K. Xu, \emph{et al}., 
Phys. Rev. Lett. \textbf{122}, 080501 (2019).

\bibitem{zhu2019}
Z. Zhu, T. Chen, X. Yang, J. Bian, Z.-Y. Xue, and X. Peng,
Phys. Rev. Appl. \textbf{12}, 024024 (2019).


\bibitem{yinyi}
Z. Zhang, P. Z. Zhao, T. Wang, L. Xiang, Z. Jia, P. Duan, D. M. Tong, Y. Yin, and G. Guo,
New J. Phys. \textbf{21}, 073024 (2019).

\bibitem{aimz}
M.-Z. Ai, S. Li, Z. Hou, R. He, Z.-H. Qian, Z.-Y. Xue, J.-M. Cui, Y.-F. Huang, C.-F. Li, and G.-C. Guo,
Phys. Rev. Appl. \textbf{14}, 054062 (2020).


\bibitem{yuyang} Z. Han, Y. Dong, B. Liu, \emph{et al}., 
arXiv:2004.10364 (2020).

\bibitem{aimz2}
M.-Z. Ai, S. Li, R. He, Z.-Y. Xue, J.-M. Cui, Y.-F. Huang, C.-F. Li, and G.-C. Guo,
arXiv:2101.07483.


\bibitem{sunfw1} Y. Dong, S.-C. Zhang, Y. Zheng, H.-B. Lin, L.-K. Shan, X.-D. Chen, W. Zhu, G.-Z. Wang, G.-C. Guo, and F.-W. Sun,
 arXiv:2102.09227.



\bibitem{sunfw2} Y. Dong, C. Feng, Y. Zheng,  X.-D. Chen, G.-C. Guo, and F.-W. Sun,
arXiv:2105.05481.

\bibitem{lisai}  S. Li, B.-J. Liu, Z. Ni, \emph{et al}., 
arXiv:2106.03474.




\bibitem{dc1} K. Khodjasteh and L. Viola,
Phys. Rev. Lett. \textbf{102}, 080501 (2009).


\bibitem{dc2}  X. Wang, L. S. Bishop, J. P. Kestner, E. Barnes, K. Sun, and S. D. Sarma,
Nat. Commun. \textbf{3}, 997 (2012).


\bibitem{dc3}
X. Rong, J. Geng, F. Shi, Y. Liu, K. Xu, W. Ma, F. Kong, Z. Jiang, Y. Wu, and J. Du,
Nat. Commun. \textbf{6}, 8748 (2015).


\bibitem{cqed} A. Wallraff, D. I. Schuster, A. Blais, \emph{et al}., 
Nature (London) \textbf{431}, 162 (2004).


\bibitem{wangym1}
Y. Wang, J. Zhang, C. Wu, J. Q. You, and G. Romero,
 Phys. Rev. A \textbf{94}, 012328 (2016).

 \bibitem{xue}
Z.-Y. Xue, F.-L. Gu, Z.-P. Hong, Z.-H. Yang, D.-W. Zhang, Y. Hu, and J. Q. You,
Phys. Rev. Appl. \textbf{7}, 054022 (2017).

\bibitem{wangym}
Y. Wang, Y. Su, X. Chen, and C. Wu,
 Phys. Rev. Appl. \textbf{14}, 044043 (2020).




\bibitem{Polariton}
Z.-Y. Xue, W.-C. Yu, L.-N. Yang, and Y. Hu,
Eur. Phys. J. D \textbf{69}, 57 (2015).

\bibitem{cavity} M. Reagor, H. Paik, G. Catelani \emph{et al}., 
Appl. Phys. Lett. \textbf{102}, 192604 (2013).

\bibitem{tunephoto}  M. C. Collodo, A. Poto\v{c}nik, S. Gasparinetti, J.-C. Besse, M. Pechal, M. Sameti, M. J. Hartmann, A. Wallraff, and C. Eichler,
Phys. Rev. Lett. \textbf{122}, 183601  (2019).


\bibitem{gatefidelity}
J. F. Poyatos, J. I. Cirac, and P. Zoller,
Phys. Rev. Lett. \textbf{78}, 390 (1997).

}
\end{thebibliography}
\end{document}